\documentclass[
twocolumn,
amsmath,
amssymb,
floatfix,
aps,
prl,
longbibliography
]{revtex4-1}

\usepackage[english]{babel}
\usepackage{hyperref}
\usepackage[T1]{fontenc}
\usepackage[utf8]{inputenc}
\usepackage{textcomp} 
\usepackage{lmodern} 
\usepackage{amsmath}
\usepackage{xcolor}
\usepackage{gensymb}

\usepackage{txfonts}
\usepackage{amsfonts}
\usepackage{graphicx}
\graphicspath{{./figures/final/}}

\hypersetup{
    breaklinks = true,
    colorlinks = true,
    pdftitle = {}, 
    pdfauthor = {},
    pdfkeywords = {},
    linkcolor = blue,
    citecolor = blue,
    filecolor = black,
    urlcolor = blue
}

\newcommand{\feco}{Fe$_{1-x}$Co$_{x}$}
\newcommand{\tc}{$T_{\rm C}$}

\newcommand{\muevat}{$\muup$eV\,atom$^{-1}$}

\newcommand{\mubat}{$\muup_{\rm B}$\,atom$^{-1}$}
\newcommand{\mjmqb}{MJ\,m$^{-3}$}

\newcommand{\etal}{\textit{et~al.}}
\newcommand{\bctten}{bct$_{10}$}

\definecolor{hlcolor}{RGB}{225, 25, 0}

\begin{document}
\begin{sloppypar} 

%
\title{Giant magnetocrystalline anisotropy energy in Fe--Co alloy under uniaxial compression: first-principles prediction}

\author{Wojciech Marciniak}%
	\email[email: ]{wojciech.marciniak@put.poznan.pl}%
	\affiliation{Institute of Molecular Physics, Polish Academy of Sciences,  M. Smoluchowskiego 17, 60-179 Pozna\'n, Poland}%
	\affiliation{Institute of Physics, Poznan University of Technology, Piotrowo 3, 60-965 Pozna\'n, Poland}%
 \author{Jos\'e \'Angel Castellanos-Reyes}%
	\affiliation{Department of Physics and Astronomy, Uppsala University, Box 516, 75120 Uppsala, Sweden}%
  \author{Joanna Marciniak}%
	\affiliation{Institute of Molecular Physics, Polish Academy of Sciences,  M. Smoluchowskiego 17, 60-179 Pozna\'n, Poland}%
\author{Miros{\l}aw Werwiński}%
	\affiliation{Institute of Molecular Physics, Polish Academy of Sciences,  M. Smoluchowskiego 17, 60-179 Pozna\'n, Poland}%

\begin{abstract}

Uniaxially strained Fe\textendash{}Co disordered alloys have emerged as promising candidates for cost-effective rare-earth-free permanent magnets due to their high magnetocrystalline anisotropy energy (MAE). 
Using first-principles, fully relativistic calculations within the coherent potential approximation and PBE exchange-correlation potential, we explore the MAE of tetragonal Fe\textendash{}Co alloys under uniaxial compression. 
Our results reveal a previously uncharted high-MAE (exceeding 3\,\mjmqb{}) region, distinct from known structures and accessible through uniaxial compression. 
%

\end{abstract}


\maketitle


\begin{figure*}
    \centering
    \includegraphics[width=0.95\textwidth]{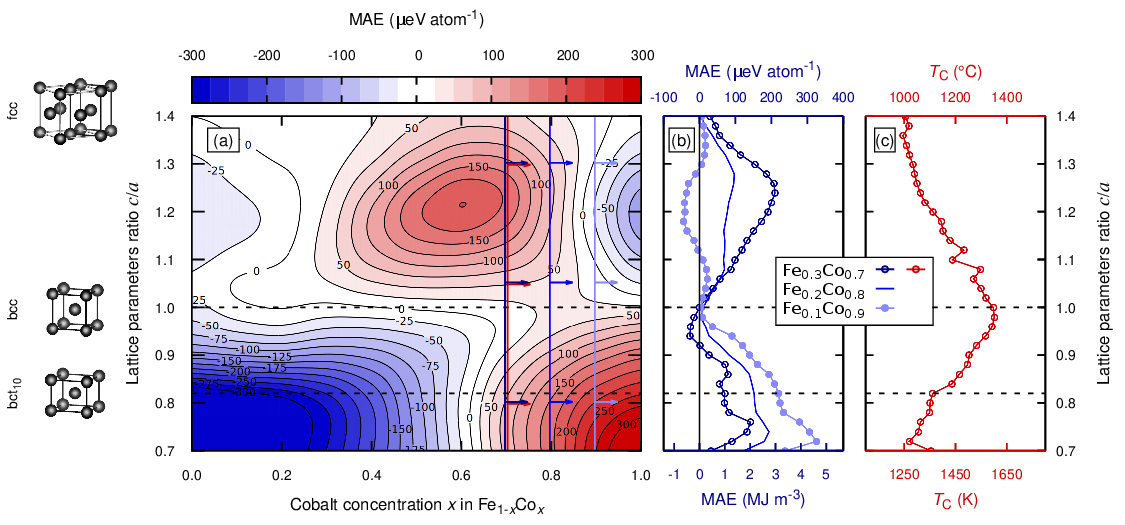}
    \caption{\label{fig:mae-map} Magnetic properties of \feco{} alloys under tetragonal deformation. Icons on the left present the structure distortion, with light gray lines in the fcc structure depicting the conventional, 4-atom unit cell. Panel (a) presents a Gaussian-smoothed map of the magnetocrystalline anisotropy energy (MAE) calculated in the coherent potential approximation (CPA) of chemical disorder treatment. Panel (b) displays the raw, non-approximated data for three different cobalt concentrations: $x=0.7$ (dark blue, empty circles), $x=0.8$ (blue, solid line), and $x=0.9$ (light blue, filled circles). Panel (c) shows the Curie temperature (\tc{}) as obtained from Monte Carlo simulations for $x=0.7$.
    }
\end{figure*}

%
The 2011 \textit{rare-earth crisis} highlighted the vulnerability of the rare-earth (RE) elements market and spurred efforts to identify cost-effective, RE-free permanent magnets~\cite{skomski_permanent_2016}. 
This included a thorough revision of ideas proposed in the mid and late 20th century~\cite{coey_perspective_2020,skomski_magnetic_2016}. 
%
%
In particular, disordered \feco{} alloys with nearly 60~at\% of cobalt and a uniaxial tetragonal strain close to 22\% has emerged as a promising candidate for a RE-free permanent magnet already in 2004.
Virtual crystal approximation (VCA) calculations predict for this alloy exceptional magnetic properties, including a giant magnetocrystalline anisotropy energy (MAE) of~$\sim$800~\muevat{}, a significant magnetic moment of 2.5~\mubat{}~\cite{burkert_giant_2004}, and a high Curie temperature (\tc{}) exceeding 1000$\degree$C~\cite{lezaic_first-principles_2007,hasegawa_challenges_2021}.


The synthesis of epitaxial films along Bain's path~\cite{alippi_strained_1997} has been proposed as a viable fabrication route for this alloy~\cite{burkert_giant_2004}, resulting in several experimental realizations in years 2006\textendash{}2009~\cite{winkelmann_perpendicular_2006,yildiz_volume_2009,yildiz_strong_2009,luo_tuning_2007}. 
%
%
At the same time, progress was made towards obtaining bulk structures~\cite{andersson_perpendicular_2006,warnicke_magnetic_2007}.
Another advance was the successful epitaxial growth of FeCo on AuCu substrate~\cite{giannopoulos_large_2015,gong_phase_2014}. 
%
%
Unfortunately, both ordered and disordered Fe\textendash{}Co alloys rapidly relax into the cubic bcc structure beyond a critical thickness of about 15 atomic monolayers ($\sim$2 nm)~\cite{reichel_increased_2014}.
This abrupt structural relaxation can be partly mitigated by interstitial doping with smaller atoms such as B, C, or N \textendash{} as supported by both theoretical predictions~\cite{delczeg-czirjak_stabilization_2014,marciniak_structural_2023} and experimental evidence~\cite{giannopoulos_large_2015,giannopoulos_coherently_2018,reichel_soft_2015,reichel_lattice_2015}. 


Further theoretical studies, conducted between 2011 and 2016, explain this relaxation and lower-than-expected measured MAE values. 
It has been shown that VCA applied in Ref.~\cite{burkert_giant_2004} tends to overestimate the MAE in Fe\textendash{}Co systems. 
More advanced chemical disorder treatment methods have demonstrated a significant reduction in MAE, with values approaching 200~\muevat{} ($\sim$2.5~\mjmqb{}) for an optimal lattice parameters $c/a$ ratio of 1.22~\cite{turek_magnetic_2012,delczeg-czirjak_stabilization_2014,steiner_calculation_2016,marciniak_structural_2023}. 


The distinctive magnetic properties of Fe\textendash{}Co alloy in the body-centered tetragonal (bct) structure, with $c/a$ ratio close to 1.22, are attributed to symmetry changes in hybridized iron and cobalt orbitals~\cite{burkert_giant_2004}.
Moreover, the magnetic behavior of these alloys is primarily driven by the strong Fe\textendash{}Co exchange interactions~\cite{odkhuu_first-principles_2019}.
Since other symmetry changes can similarly alter the chemical environment, it is also essential to investigate the magnetic properties of Fe\textendash{}Co alloys along alternative lattice transformation pathways.

Thus, we complement the transformation along Bain’s path with a particular bcc~$\leftrightarrow$~hcp martensitic phase transition pathway, first described by Burgers~\cite{burgers_process_1934}.
This process involves two steps: uniaxial compression along one of the main crystallographic axes, followed by the perpendicular slip of one of the two hexagonal atomic planes formed during such deformation.
In general, incorporating hexagonal cell shear in addition to the plane slip allows the description of broader fcc~$\leftrightarrow$~hcp~$\leftrightarrow$~bcc~$\leftrightarrow$~fcc transitions~\cite{ekman_ab_1998,wang_iron_1998,liu_bcc--hcp_2009}.
Simultaneous uniaxial deformation and shear (or atomic plane slip) is, however, currently considered the more likely transformation path~\cite{lu_does_2014,friak_ab_2008}.
The described deformation continues the well-known~\cite{burkert_calculation_2004} tetragonal distortion along Bain’s path into the $c/a < 1$ regime.
%
%
The intermediate structure, for which the coordination number of the atoms changes to 10, is denoted further as \bctten{} and has been studied for various elements~\cite{mehl_absence_2004}. 
This structure forms through uniaxial deformation along the [001] direction of the tetragonal phase, transforming the bcc (110) into the hexagonal (0001) plane.


Firstly, we examined the \feco{} system under uniaxial tetragonal distortion, ranging from $c/a$\,$\approx$\,1.42 (fcc structure) through the \bctten{} phase ($c/a = \sqrt{2/3} \approx 0.82$) in the Burgers transformation, down to $c/a$\,$\approx$\,0.7. 
We employed the spin-polarized relativistic Korringa-Kohn-Rostoker method, implemented in SPR-KKR version 7.7.1~\cite{ebert_calculating_2011}.
A 2-atom basis and space group $P4/mmm$ were used to ensure consistency in basis and symmetry across all structures.
Each site was occupied evenly with the chemical disorder treated via coherent potential approximation (CPA).
%
We optimized the unit cell volume for various cobalt concentrations, $x$, in \feco{}, with a step of 10\%.
Convergence tests indicated a requirement of at least 10,000 \textbf{k}-points and an angular momentum expansion cutoff $l_{\rm max}$~=~4 to properly capture the hybridization of Fe/Co 3$d$ orbitals~\footnote{This led to the characteristic Slater-Pauling-like shape in the unit cell volume \textit{versus} $x$ dependency (optimization results not shown).}.
For the exchange-correlation potential, we used the generalized gradient approximation (GGA) with the Perdew, Burke, and Ernzerhof (PBE) parametrization~\cite{perdew_generalized_1996}, without shape approximation (full potential approach).

For the assessment of the MAE dependency on $c/a$ and cobalt concentration $x$, we employed a constant volume approach, where the lattice parameters were rescaled to conserve the computational cell volume optimized for each cobalt concentration. 
We explored the entire Bain transition path between fcc and bcc structures along the $z$-axis distortion. 
The MAE of each structure was determined by calculating the energy difference between the [001] and [100] magnetization directions as follows: 

\begin{equation}
    {\rm MAE} = E_{\rm 100} - E_{\rm 001},
\end{equation}
with the energies calculated in the four-component relativistic formalism, including the spin-orbit coupling.


The nonmagnetic global energy minimum of fcc iron has been thoroughly investigated in numerous works, including key \feco{} studies~\cite{burkert_calculation_2004,turek_magnetic_2012,steiner_calculation_2016,neise_effect_2011}. 
Here, we aim to capture the behavior of the high-spin (HS) fcc iron phase~\cite{moruzzi_ferromagnetic_1986}, which has recently gained interest in layered systems~\cite{liang_chemical_2023,meixner_magnetic_2024}.


We began by comparing the MAE of the tetragonally strained structure, shown in Fig.~\ref{fig:mae-map}(a), to the results of works by Burkert~\etal~\cite{burkert_calculation_2004}, Turek~\etal{}~\cite{turek_magnetic_2012}, Steiner~\etal{}~\cite{steiner_calculation_2016}, and Neise~\etal{}~\cite{neise_effect_2011}. 
Our calculated MAE maximum \textendash{} of about 200~\muevat{} (2.5~\mjmqb{}) for $c/a$~=~1.22 and $x$~=~0.6 \textendash{} is in good agreement with results obtained previously using CPA and explicit random alloy models. 
As expected, this value is about four times lower than the 800~\muevat{}, initially predicted by Burkert~\etal{}
The overall features, such as the observed uniaxial MAE maximum and the in-plane anisotropy region for the Co-rich system, are consistent with previous findings. 
The noticeable difference is the complete vanishing of the MAE when considering the fcc iron HS phase, whereas prior studies reported a minor residual MAE.
Importantly, we uncover a second, prominent MAE maximum for \feco{} under strong uniaxial compression, spanning a wide range of $x$ for the Co-rich structures. 
MAE value near this maximum easily exceeds 3\,\mjmqb{} and is an order of magnitude higher than that of the bulk hcp cobalt~\cite{aas_effect_2013,daalderop_firstprinciples_1990a}.


Qualitatively, the mutual part of our results (stoichiometric FeCo) aligns with the recent findings of Wolloch and Suess~\cite{wolloch_strain-induced_2021}. 
However, quantitative comparisons are limited due to our exclusion of surface effects and their use of an insubstantial quantity of special quasirandom structures models.
%
%
%
%
%
In Fig.~\ref{fig:mae-map}(b), we see that the MAE increases sharply with $c/a$ decreasing to around 0.85 and remains stable during further compression down to around $c/a$ = 0.78.
Afterwards, it increases sharply even more.
Unfortunately, this extreme compression is accompanied by a rapid growth in system energy, as we will demonstrate later.


We find that for cobalt concentrations around 0.75, high uniaxial MAE in the strained structure is accompanied by similarly high MAE during equivalent compression. 
Hence, when the distance between neighboring atoms increases \textendash{} due to, e.g., interstitial doping or finite-temperature lattice vibrations \textendash{} the distance between those atoms and their respective neighbors on the other side decreases, further increasing MAE. 
This explains the appearance of a broader MAE maximum, shifted toward higher cobalt concentrations, as observed in studies using various configurational space sampling methods in B- and C-doped Fe\textendash{}Co~\cite{delczeg-czirjak_stabilization_2014, marciniak_structural_2023, neise_effect_2011}. 
Notably, phonon-induced magnetism enhancement has already been noted in cobalt-doped iron garnets~\cite{frej_phonon-induced_2023}.


Based on the self-consistently converged electron density, for $x$ = 0.7, we calculated exchange integrals $J_{\rm ij}$ up to the 9$^{\rm th}$ neighboring cell using the Lichtenstein~\etal{} method~\cite{liechtenstein_exchange_1984} implemented in SPR-KKR package.
Those were subsequently used for Monte Carlo simulations in the Uppsala atomistic spin dynamics (UppASD) package~\cite{eriksson_atomistic_2017}.
The Curie temperature was determined using the Binder cumulant method, incorporating finite-size scaling~\cite{eriksson_atomistic_2017,landau_guide_2021,noauthor_tutorial_nodate}.

%
Figure~\ref{fig:mae-map}(c) shows the estimated \tc{} values derived from the intersection of the Binder cumulant \textit{versus} temperature curves in two supercells \textendash{} sized $25^3$ and $30^3$ \textendash{} based on the methodology outlined in Ref.~\cite{noauthor_tutorial_nodate}.
Our calculations suggest that for the \bctten{} structure, the expected \tc{} ranges between 1250 and 1350 K, a highly suitable value for permanent magnet applications. 
These estimates align with prior results from the group of Bl\"ugel using mean-field theory~\cite{lezaic_first-principles_2007,jakobsson_tuning_2013}, which reported \tc{} values of 1550~K for bcc Fe$_{0.3}$Co$_{0.7}$ and 1520~K for bcc Fe$_{0.25}$Co$_{0.75}$, with an approximate 20\% \tc{} reduction under uniaxial compression. 
Additionally, we observe that the \tc{} \textit{versus} $c/a$ dependence is slightly asymmetric, favoring uniaxial compression over uniaxial strain.


\begin{figure}
    \centering
    \includegraphics[width=\columnwidth]{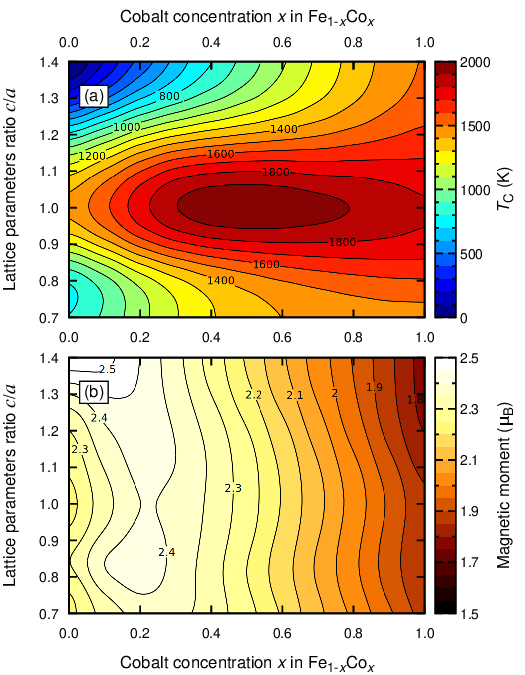}
    \caption{\label{fig:mom-tc-maps} (a) Curie temperature (\tc{}) \textit{versus} $c/a$ lattice parameters ratio and cobalt concentration $x$ for the \feco{} alloy under tetragonal deformation. (b) Corresponding total (spin $+$ orbital) magnetic moment map. Calculations were performed using the SPR-KKR 7.7.1 package, employing CPA for chemical disorder and the PBE exchange-correlation potential. \tc{} was estimated using the disordered local moments (DLM) method.}
\end{figure}


Due to computational limitations, we were restricted in performing high-accuracy Monte Carlo \tc{} calculations. 
Therefore, we employed the disordered local moments (DLM) method to derive a qualitative map of the critical temperature across the same $c/a$ and $x$ ranges. 
This method approximates the material's paramagnetic state as a CPA-like state, where each atomic site hosts mixed antiparallel magnetic moments (DLM state). 
The critical temperature of the ferromagnetic system can then be related to thermal energy and estimated approximately as follows~\cite{gyorffy_first-principles_1985,bergqvist_theoretical_2007}:

\begin{equation}
    T_{\rm critical}^{\rm DLM} = \frac{2}{3}\frac{E_{\rm DLM} - E_{\rm FM}}{k_{\rm B} \cdot c},
\end{equation}
if the ferromagnetic state is energetically favorable, and:

\begin{equation}
    T_{\rm critical}^{\rm DLM} = 0
\end{equation}
otherwise, where $E_{\rm DLM}$ and $E_{\rm FM}$ are the energies of the DLM and ferromagnetic states, respectively, $c$ is the number of magnetic atoms in the system, and $k_{\rm B}$ is the Boltzmann constant.
In Fig.~\ref{fig:mom-tc-maps}, we present a calculated $T_{\rm C}^{\rm DLM}$ map alongside the corresponding magnetic moments map.


Since the DLM state is susceptible to a collapse of the magnetic moment, we could not obtain a stable paramagnetic HS state. 
Consequently, as shown in Fig.~\ref{fig:mom-tc-maps}(a), \tc{} tends toward zero when approaching the iron fcc phase. 
In contrast, as depicted in Fig.~\ref{fig:mom-tc-maps}(b), the ferromagnetic HS phase was successfully reproduced, with a magnetic moment value of 2.5~\mubat{}, consistent with previous studies~\cite{moruzzi_ferromagnetic_1986,meixner_magnetic_2024}.
%
%
A key feature of the \tc{} and magnetic moment maps is the presence of Slater-Pauling-like maxima for cubic structures, observed at $x \approx 0.5$ for the \tc{} and at $x \approx0.2 $ for the magnetic moment. 
%
%
Both maps show promising results, with slightly asymmetric dependencies that favor the compressed structure across all cobalt concentrations. 
However, weak minima are observed in the magnetic moment \textit{versus} $c/a$ dependency at $c/a \approx 0.82$ and cobalt concentrations above $x=0.4$.
Given that the DLM method typically overestimates the result by approximately 30\%~\cite{ebert_calculating_2011,ke_effects_2013,hedlund_magnetic_2017}, we estimate that \tc{} for Co-rich \bctten{} structures should lie between 1250 and 1350 K. 
We also expect the average magnetic moment to remain considerably high, between 1.8 and 2.2~\mubat{}. 
After accounting for the expected overestimation by the DLM method, our results agree with both: our Monte Carlo calculations (Fig.~\ref{fig:mae-map}(c)) and previous work by the group of Bl\"ugel~\cite{lezaic_first-principles_2007,jakobsson_tuning_2013}. 
Specifically, our findings align with the high \tc{} values predicted for body-centered tetragonal cobalt~\cite{lezaic_first-principles_2007}.


\begin{figure}
    \centering
    \includegraphics[width=\columnwidth]{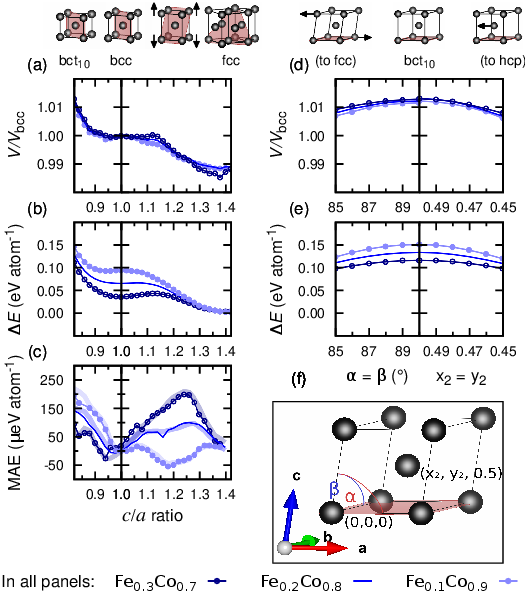}
    \caption{\label{fig:hcp-bct-fcc-transition} Panels (a) and (d) depict the optimized cell volume ratio relative to the bcc structure. Panels (b) and (e) show the stability of the distorted \feco{} alloy expressed in the form of transformation energy. Panel (c) presents the influence of geometry optimization on the MAE. The colors correspond to those in Fig.~\ref{fig:mae-map}, with additional thick solid lines representing the unoptimized structure and thin lines with circles representing volume-optimized structures. Panels (a--c) depict the tetragonal structure under uniaxial distortion along the Bain path and beyond. Panels (d) and (e) show the skew and slip of hexagonal planes along the Burgers path. Panel (f) presents the $P1$ structure parameters that vary during the Burgers transformation. The atomic plane marked in red in the miniatures denotes the hexagonal plane: (101) in the \bctten{} structure, (111) in the fcc structure realized in $Pm\bar{3}m$ space group, and (001) in the structures along the Burgers path, realized in the $P1$ space group.}
\end{figure}

Two significant concerns arise. 
Firstly, the applied constant volume approximation may introduce certain distortions in the picture of such intricate value as MAE.
Secondly, developing a technological process to stabilize the \bctten{} structure may be possible, potentially making it easier to achieve experimentally than the structure in the MAE maximum proposed by Burkert~\etal{}
%
%
Despite that the optimal cobalt structure is hcp, over 30\,nm thick layers of bcc cobalt have been obtained~\cite{prinz_stabilization_1985} and tetragonal structure proven to be more stable~\cite{giordano_evidence_1996}.
Hence, in the last part of this work, we will focus on assessing the approximation accuracy and the structure's stability.

To optimize the unit cell volume along the bcc~$\leftrightarrow$~fcc~$\leftrightarrow$~hcp transformation path, we used the full-potential local-orbital (FPLO) code~\cite{koepernik_full-potential_1999,eschrig_chapter_2004}, version 5.00-18.
We applied the local density approximation (LDA) of exchange-correlation potential, based on the Perdew and Wang (PW92) parametrization~\cite{perdew_accurate_1992}, and used $P1$ symmetry. 
We sacrificed the accuracy of the PBE potential and fully relativistic approach \textendash{} excessive for the geometry optimization \textendash{} to adhere to the presumption of using CPA. 
Although exact volume values differ between the LDA and GGA, we assumed that the volume ratios between any two structures would remain unchanged. 
Thus, we used the $V/V_{bcc}$ profile to rescale the system volume for transformation energy ($\Delta E$) and MAE calculations, which were carried out using the SPR-KKR/PBE framework described above. 
The results are presented in Fig.~\ref{fig:hcp-bct-fcc-transition}.


As expected, the unit cell volume decreases when approaching the fcc and hcp (close-packed) structures. 
Overall, the optimized unit cell volume (Figs.~\ref{fig:hcp-bct-fcc-transition}(a) and \ref{fig:hcp-bct-fcc-transition}(d)) remains within $\pm$1.25\% of the bcc unit cell volume.
This is slightly lower than the $\sim$2\% value predicted by experiments, but consistent with earlier density functional theory calculations~\cite{marcus_equilibrium_1985,moruzzi_ferromagnetic_1986}.
%
%
Our results indicate that the constant volume approximation does not impact the MAE significantly. 


The energy profile along the bcc~~$\leftrightarrow$~~\bctten{} path (Fig.~\ref{fig:hcp-bct-fcc-transition}(b)) reveals a shallow minimum, which forms when $x > 0.8$. 
At the same time, along the fcc~~$\leftrightarrow$~~\bctten{}~~$\leftrightarrow$~~hcp transition path (Fig.~\ref{fig:hcp-bct-fcc-transition}(e)), the \bctten{} structure exhibits a local energy maximum.
The two results taken together indicate a saddle point on the energy surface. 
While the minimum on the bcc~~$\leftrightarrow$~~\bctten{} path is relatively shallow and close to more pronounced minima, we expect that the \bctten{} structure could be stabilized by combining interstitial doping with epitaxial growth on a suitable square lattice \textendash{} a solution investigated by Mehl~\etal{}~\cite{mehl_absence_2004}.
It is primarily because the $\Delta E$ presented in Fig.~\ref{fig:hcp-bct-fcc-transition}(e) is almost flat in a wide range of transformations.
We also expect the MAE in such epitaxial thin films \textendash{} especially made of pure Co \textendash{} to grow even further~\cite{daalderop_firstprinciples_1990a}.


In summary, we present a comprehensive study of magnetocrystalline anisotropy energies (MAE), magnetic moments, and Curie temperatures (\tc{}) of Fe\textendash{}Co alloys across a wide range of tetragonal deformations.
We also complete the investigation of the stability of the alloy by bridging the Bain and Burgers transformation pathways. 
While using state-of-the-art computational methods, this work covers a broader range of deformations than has been explored before.
To validate the employed calculation techniques, we first reproduced key results from earlier studies, including the MAE and magnetic moment maps for body-centered tetragonal (bct) structures with $c/a>1$, recreating the well-known Burkert~\etal{} maximum in MAE. 
We then extended our study to Fe\textendash{}Co alloys under uniaxial stress, a region that has rarely been investigated.

Our findings reveal a high-MAE region in uniaxially compressed Fe\textendash{}Co alloys, particularly for Co-rich alloys. 
This region centers nearby an intermediate \bctten{} structure within the proposed Burgers bcc~$\leftrightarrow$~hcp~$\leftrightarrow$~fcc transition pathway.
%
%
The new structure is located at a saddle point of the energy surface. 
Consequently, it could be stabilized \textendash{} e.g., by epitaxial growth on suitable square-lattice substrates or/and by interstitial doping. 
This could lead to stable structures with much thicker layers than those previously reported for uniaxially strained Fe\textendash{}Co alloys.
Monte Carlo simulations and DLM calculations indicate that this Co-rich structure under uniaxial tetragonal compression exhibits a higher \tc{} than under strain. 
Ultimately, our work broadens the range of stable, high-MAE structures available for various cobalt contents in tetragonal Fe\textendash{}Co alloys. 
This now spans from ordered systems with cobalt concentrations as low as 25\%, through stoichiometric FeCo, to nearly pure cobalt \textendash{} as identified in this study.

\section*{Acknowledgements}

\begin{acknowledgments}

We acknowledge the financial support of the National Science Centre Poland under the decision DEC-2018/30/E/ST3/00267. 
J.~Á.~C.-R. acknowledges the financial support from the Olle Engkvist Foundation and the Knut and Alice Wallenberg Foundation. 
The collaboration between W.~M., J.~M., and J.~Á.~C.-R. was possible thanks to the financial support of the Polish National Agency for Academic Exchange under the decision BPN/BEK/2022/1/00179/DEC/1.
The main computational work was performed at the Poznan Supercomputing and Networking Centre (PSNC/PCSS).
We are grateful to Paweł Leśniak and Daniel Depcik for their assistance with compiling and maintaining the computational codes.
\tc{} calculations were enabled by resources provided by the National Academic Infrastructure for Supercomputing in Sweden (NAISS) at NSC Centre, partially funded by the Swedish Research Council through grant agreement no.\ 2022-06725.
We thank Jan Rusz for his valuable suggestions and discussions.
We also thank Mateusz Kowacz, Zbigniew Śniadecki, Karol Synoradzki, and Justyn Snarski-Adamski for their comments on the manuscript.
We thank Jan Martinek, whose insightful question inspired us to conduct this study.

\end{acknowledgments}

\end{sloppypar}

\bibliography{FeCo}

\end{document}